\documentclass[manuscript]{emulateapj}

\usepackage{graphicx}

\shorttitle{PeV neutrinos from BL Lacs}
\shortauthors{Tavecchio et al.}

\begin{document}


\title{Structured jets in BL Lac objects: efficient PeV neutrino factories?}


\author{Fabrizio Tavecchio, Gabriele Ghisellini}
\affil{INAF -- Osservatorio Astronomico di Brera, Via E. Bianchi 46, I--23807 Merate, Italy}

\author{Dafne Guetta}
\affil{INAF--Osservatorio astronomico di Roma, via Frascati 33, 00040 Monte Porzio Catone, Italy}
\affil{Department of Physics Optical Engineering, ORT Braude, P.O. Box 78, Carmiel, Israel}



\begin{abstract}
The origin of the high-energy neutrinos (0.1$-$1 PeV range) detected by IceCube remains a mystery. In this work we explore the possibility that efficient neutrino production can occur in structured jets of BL Lac objects, characterized by a fast inner spine surrounded by a slower layer. This scenario has been widely discussed in the framework of the high-energy emission models for BL Lacs and radiogalaxies. One of the relevant consequences of a velocity structure is the enhancement of the inverse Compton emission caused by the radiative coupling of the two zones. We show that a similar boosting could occur for the neutrino output of the spine through the photo-meson reaction of high-energy protons scattering off the amplified soft target photon field of the layer. Assuming the local density and the cosmological evolution of $\gamma$-ray  BL Lac  derived from {\it Fermi}-LAT data, we calculate the expected diffuse neutrino intensity, that can match the IceCube data for a reasonable  choice of the parameters.
\end{abstract}


\keywords{BL Lacertae objects: general -- neutrinos -- gamma rays: galaxies}


\section{Introduction}

The origin of  high-energy neutrinos with energies 100 TeV- few PeV detected by IceCube (Aartsen et al. 2013, 2014) remains a mystery (see e.g. Anchordoqui et al. 2014 for a review). The flux level is very close to that expected from the emission by photo-meson reactions in optically-thin UHECR ($E>10^{19}$ eV) sources (Waxman \& Bahcall 1998), but the energies of the neutrinos link them to parent cosmic rays (CR) with much smaller energies $E\sim 10^{16-17}$ eV. The measurements are consistent with an isotropic, $\nu_e:\nu_{\mu}:\nu_{\tau}=1:1:1$ flux with slope $E^{-2}$, but the absence of events above 2 PeV and, in particular, at the Glashow resonance  for $\nu_e$ at 6.3 PeV,  compellingly suggests a break or a cutoff in the spectrum close  to 1 PeV (e.g. Anchordoqui et al. 2014). 

Among all possible astrophysical sources of high-energy neutrinos, radio-loud active galactic nuclei have been widely considered in the past. In particular, the attention has generally been focused to {\it blazars}, i.e. those whose relativistic jets point toward the Earth. These sources dominate the high-energy $\gamma$-ray sky, both at GeV and TeV energies. Their powerful, relativistically boosted, non-thermal continuum ranging from the radio band to $\gamma$-ray energies, is produced within the jet by ultra-relativistic particles.
The most popular scenario assumes that the emission is entirely due to leptons through synchrotron and inverse Compton (IC) mechanisms (e.g. Ghisellini et al. 1998). Alternatively, hadronic models postulate that the high-energy emission originates from ultra-high energy hadrons, emitting through synchrotron or photo-meson mechanisms (e.g. Muecke et al. 2003). Even if hadrons are not the dominant source of high-energy radiation,  their interaction with the radiation fields naturally results in the emission of neutrinos.

Recently, Murase et al. (2014) and Dermer et al. (2014) performed a thorough analysis of the expected neutrino emission through photo-meson reactions in blazar jets under the assumption that an important CR component exists in the jet of all kind of blazars. Their results show that neutrino output is dominated by the most powerful blazars (flat spectrum radio quasar, FSRQ) with a marginal contribution by the weak blazars, the so-called BL Lac objects. Besides the low intrinsic power of the jet, the inefficient neutrino production by BL Lac is a direct consequence of the small radiation energy density in these jets. The integrated neutrino spectrum of Murase et al. (2014), however, is expected to have a maximum - strictly linked to the peak frequency of the soft target radiation field - occurring at relatively high energies, above 10 PeV. Murase et al. (2014) concluded that it is difficult to reproduce the IceCube results with the simplest emission model, without invoking complications to the standard scenario.

On the observational side, an interesting clue has been recently provided by Padovani \& Resconi (2014) who, through a correlation analysis of IceCube events and gamma-ray sources, find a suggestive positional correlation of some events with few ``classical" TeV BL Lacs, most notably  Mkn 421, PG 1553+113 and H 2356-309 (other events seem instead correlated to pulsar wind nebulae in the Milky Way). 

Motivated by these hints, we reconsider here the possible production of neutrinos in BL Lac jets. The aforementioned analysis by Murase et al. (2014) and Dermer et al. (2014) is based on the standard {\it one-zone} emission framework, assuming that a single active sub-region of the jet is responsible for the bulk of the radiation that we observe from blazars. However, there is growing evidence that the emission occurs in more complex regions. In particular, the modeling of the emission of TeV emitting BL Lacs and low power radio-galaxies (thought to be the misaligned parent population of BL Lacs, e.g. Urry \& Padovani 1995) led to postulate the existence of a structure for the jet, with a faster core (the spine) surrounded by a slower layer (Ghisellini et al. 2005, Tavecchio et al. 2008). Such a spine-layer structure is actually directly inferred from VLBI observations in the radio band, showing a ``limb-brightened" structure of some jets, whose simplest explanation is a transverse velocity structure of the jet (e.g. Giroletti et al. 2008). The basic idea of the spine-layer model is that such a structure naturally implies the enhancement of the radiative output of both components. In fact, thanks to the relativistic amplification induced by the relative motion, the radiation produced by one component can dominate the radiation field in the frame of the other, leading to an overall increased efficiency of the IC emission with respect to that of the one-zone model.  Clearly, the same principle can be  applied to the production of neutrinos, whose emission through photo-meson production by high-energy protons in the spine can be boosted by the amplification of the layer radiation field in the spine frame. Although all BL Lac objects could be characterized by a spine-layer structure, we focus here only to the so-called highly-peaked BL Lac objects (HBL, the majority of the TeV emitting BL Lacs), for which the spine-layer structure has been directly observed. As discussed by Murase et al. (2012), protons can be in principle accelerated in these jets up to maximal energies of $E\sim 10^{19}$ eV, much more than those required to produce PeV neutrino energies.

Given the exploratory nature of this work, we do not try a completely self-consistent modelization of the photon and neutrino emission. Rather, we adopt a template for the spectrum of the layer inspired by the observed spectral energy distribution of HBL and previous application of the structured jet model. We leave a more detailed study to a future work. 

 Throughout the paper, we assume a cosmology with  $H_0=70$ km s$^{-1}$ Mpc$^{-1}$, $\Omega_{\rm M}=0.3$, $\Omega_{\Lambda}=0.7$.

\section{The model}

\subsection{Structured BL Lac jets}

We briefly recall the basic features of the structured jet framework of Ghisellini et al. (2005) relevant for the present application. 

The jet is modeled as a two-fluid flow, with a spine with bulk Lorentz factor $\Gamma_{\rm s}$ and a outer layer with bulk Lorentz factor $\Gamma_{\rm l}<\Gamma_{\rm s}$. Observing the jet at a viewing angle $\theta_{\rm v}$, the spine and the layer are characterized by a relativistic Doppler factor $\delta_{\rm s,l}=[\Gamma_{\rm s,l}(1-\beta_{\rm s,l}\cos \theta_{\rm v})]^{-1}$. $Q$, $Q^{\prime}$ and $Q^{\prime\prime}$ indicate quantities measured in the observer, spine and layer reference frame.

The (soft) target radiation field of the layer in the observer frame  is parameterized by a smoothed broken power law function:
\begin{equation}
L(\epsilon_{\rm t})=k\left(\frac{\epsilon_{\rm t}}{\epsilon_{\rm o}}\right)^{-\alpha_1}\left[1 + \left( \frac{\epsilon_{\rm t}}{\epsilon_{\rm o}}\right)\right]^{\alpha_1-\alpha_2},
\end{equation}
whose normalization is given by the total {\it observed} luminosity:
$L_{\rm t}=\int L(\epsilon_{\rm t}) d\epsilon_{\rm t}$.
The corresponding photon number density in the layer frame is:
\begin{equation}
n^{\prime\prime}(\epsilon^ {\prime\prime}_{\rm t})=\frac{L(\epsilon_{\rm t})} {4\pi R^2c\delta_{\rm l}^3 \, \epsilon^ {\prime\prime}_{\rm t}},
\end{equation}
where $\epsilon^ {\prime\prime}_{\rm t}  = \epsilon_{\rm t}/\delta_{\rm l}$ and $R$ is the jet radius .

The relative motion of the two components leads to the amplification of the radiation field of the layer as observed in the spine reference frame (and {\it viceversa}). Specifically, given the number density of the soft radiation in the layer frame, the photon density in the spine  frame is $n^{\prime}_{\rm t}(\epsilon^ {\prime}_{\rm t})d\epsilon^{\prime}_{\rm t}=\Gamma_{\rm rel} \, n_{\rm t}^{\prime\prime}(\epsilon_{\rm t} ^{\prime\prime})d\epsilon_{\rm t}^{\prime\prime}$, where $\Gamma_{\rm rel} =\Gamma_{\rm s}\Gamma_{\rm l}(1-\beta_{\rm s}\beta_{\rm l})$ is the relative Lorentz factor and $\epsilon_{\rm t}^{\prime}\simeq\Gamma_{\rm rel}\epsilon_{\rm t}^{\prime\prime}$. Note that, while for the IC emission the relevant quantity is the energy density of the target photon field (i.e. the synchrotron radiation produced in the layer), which transforms as $\Gamma _{\rm rel}^2$, here we are interested to the {\it numerical} density, depending on $\Gamma_{\rm rel}$. 
For simplicity (as Atoyan \& Dermer 2003) we do not take into account the fact that in the spine frame the radiation field of the layer (dominating the photo-meson reactions) is  anisotropic (Dermer 1995). 
 
\subsection{Neutrino emission}

We assume that in the spine there is a population of CR (protons) whose luminosity (measured in the spine frame) is parametrized by a cut-offed power law distribution:
\begin{equation}
L^ {\prime}_{\rm p}(E_{\rm p}^ {\prime})=k_{\rm p} E_{\rm p}^{\prime \, -n}\exp\left( -\frac{E_{\rm p}^ {\prime}}{E_{\rm cut}^ {\prime}}\right) \;\;\; E_{\rm p}^ {\prime}>E_{\rm min}^ {\prime}
\end{equation} 
with total (spine frame) luminosity $L_{\rm p}^ {\prime}=\int L_{\rm p}^ {\prime}(E_{\rm p}^ {\prime}) dE_{\rm p}^ {\prime}$. 
 
The neutrino yield (in the spine frame) through the decay of the pions produced by the photo-pion reactions of protons, $\pi ^{\pm}\to \mu^{\pm}+\nu_{\mu}\to e^{\pm} + 2\nu_{\mu} +\nu_{\rm e}$  (we do not distinguish among $\nu$ and $\bar{\nu}$) is  parametrized (e.g. Murase et al. 2014) by $f_{p\gamma}(E_{\rm p}^ {\prime})=t_{\rm dyn}^ {\prime}/t_{p\gamma}^ {\prime}(E_{\rm p}^ {\prime})$, in which $t_{\rm dyn}^ {\prime}\approx R/c$ is the dynamical timescale and the (inverse of the) photo meson cooling time is given by:
\begin{equation}
t^{\prime \, -1}_{p\gamma}(E_{\rm p}^{\prime})=c \int_{\epsilon_{\rm th}} ^{\infty} d\epsilon \frac{n_{\rm t}^ {\prime}(\epsilon)}{2\gamma_{\rm p}^{\prime}\epsilon^2} \int_{\epsilon_{\rm th}}^{2\epsilon\gamma_{\rm p}^{\prime}} d{\bar\epsilon}\, \sigma_{p\gamma}({\bar\epsilon})\, K_{p\gamma}({\bar\epsilon}) \,  {\bar\epsilon},
\label{tpg}
\end{equation}
where $\gamma _{\rm p}^{\prime}=E_{\rm p}^{\prime}/m_{\rm p}c^2$, $\sigma_{p\gamma}(\epsilon)$ is the photo-pion cross section, $K_{p\gamma}(\epsilon)$ the inelasticity and $\epsilon_{\rm th}$ is the threshold energy of the process. We evaluate the integrals in Eq. \ref{tpg} using the simple but accurate prescription for $\sigma _{p\gamma}$ and $K_{p\gamma}$ provided in Atoyan \& Dermer (2003), including both single pion (from the $\Delta^+$ resonance) and multi-pion reactions. Since, as we verify below, the target radiation field in the spine reference frame is dominated by the beamed layer component, we only consider it in the integral.

The resulting neutrino luminosity in the spine frame is given by (e.g. Murase et al. 2014):
\begin{equation}
E_{\nu}^{\prime} L^{\prime}_{\nu}({E_{\nu}^{\prime}}) \simeq \frac{3}{8} f_{p\gamma}(E_{\rm p}^ {\prime}) \, E_{\rm p}^{\prime} L_{\rm p}^{\prime}({E_{\rm p}^{\prime}}); \;\;\;\; E_{\nu}^{\prime}=0.05\, E_{\rm p}^{\prime}
\end{equation}
where the factor $3/8$ takes into account the fraction of the energy going into $\nu$ and $\bar{\nu}$ (of all flavors).  For completeness we also calculate the luminosity of photons (from the $\pi^0\to 2\gamma$ decay), using the same equation and the factor $1/2$ instead of $3/8$. The contribution of the possible synchrotron emission of CR is negligible (e.g. Tavecchio et al. 2014).

Finally we calculate the neutrino (and the photon) luminosity in the observer frame using the standard transformations: $ E_{\nu} L_{\nu}({E_{\nu}}) = E_{\nu}^{\prime} L^{\prime}_{\nu}({E_{\nu}^{\prime}})\, \delta_{\rm s}^4$ and $E_{\nu}=\delta_{\rm s}E_{\nu}^{\prime}$.

\subsection{Diffuse intensity}

The procedure described in the previous section allows us to derive the neutrino output from a single BL Lac object. 
The cumulative diffuse emission from a population of BL Lacs, each one emitting a neutrino luminosity as calculated above, is determined using:
\begin{equation}
E_{\nu}I(E_{\nu})= \frac{c}{4\pi H_o} E_{\nu}^2 \int \frac{j[E_{\nu}(1+z),z]}{\sqrt{\Omega_{\rm M}(1+z)^3+\Omega_{\Lambda}}} \, dz,
\end{equation}
in which the comoving volume neutrino emissivity is given by the product of the comoving density of sources $\Sigma(z)$ and the source neutrino luminosity:
\begin{equation}
j(E_{\nu},z)=\Sigma(z) \, \frac{L_{\nu}({E_{\nu}})}{E_{\nu}}.
\end{equation}

The cosmological evolution of $\gamma$-ray emitting blazars have been recently studied by Ajello et al. (2014) using {\it Fermi}-LAT data. As already noted we will focus our calculations to HBL, the majority of the TeV emitting BL Lacs. Local HBL show a {\it negative} evolution, i.e. $\Sigma(z)$ decreases with $z$. Although complex relations are used by Ajello et al. (2014) to model the luminosity-dependent  evolution of the blazar density, here for simplicity we parametrize the evolution of the HBL as:
$\Sigma(z)=\Sigma_{\rm o}(1+z)^{-\beta}$.
From Fig. 10 of Ajello et al. (2014) one infers $\Sigma_{\rm o}\simeq 2\times 10^{-7}$ Mpc$^{-3}$ and the fast decrease of the density is reproduced by $\beta\sim 6$ (we checked that the results are only weakly dependent on the precise value of $\beta$).

A similar calculation provides the expected flux of CR from HBL, assuming that they can efficiently escape from the jets and are not substantially deviated by magnetic fields. For the energies of interest here ($E_{\rm p}<10^{18}$ eV) and for local sources such as HBL, the propagation losses are negligible (e.g., Berezinsky et al. 2006). If we allow the escape from the jet, the CR cumulative flux is limited by the observed flux at Earth. 

\begin{figure}[t]    
\centering
\includegraphics[width=.45\textwidth,height=.4\textwidth]{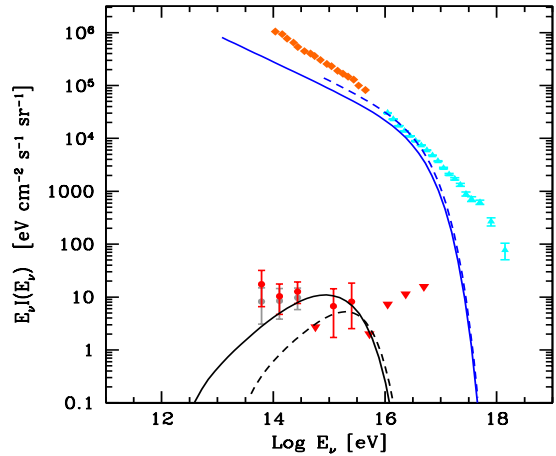}
\vspace*{-0.4 truecm}
\caption{\label{bkg} Measured diffuse intensities of high-energy neutrinos (red symbols, from Aartsen et al. 2014). Red triangles indicate upper limits. Gray data points show the fluxes for an increase of the prompt atmospheric background to the level of 90\% CL limit. Black dashed and solid lines report the diffuse neutrino intensity for model 1 and model 2, respectively. The blue lines report the corresponding CR intensities, assuming efficient escape from the jet. Orange (Apel et al. 2012) and cyan (Chen 2008) data points show the observed high-energy CR spectrum.}
\end{figure}  

\section{Results}

\begin{figure}[t]    
\centering
\hspace*{-2.1truecm}
\includegraphics[width=.7\textwidth,height=.58\textwidth]{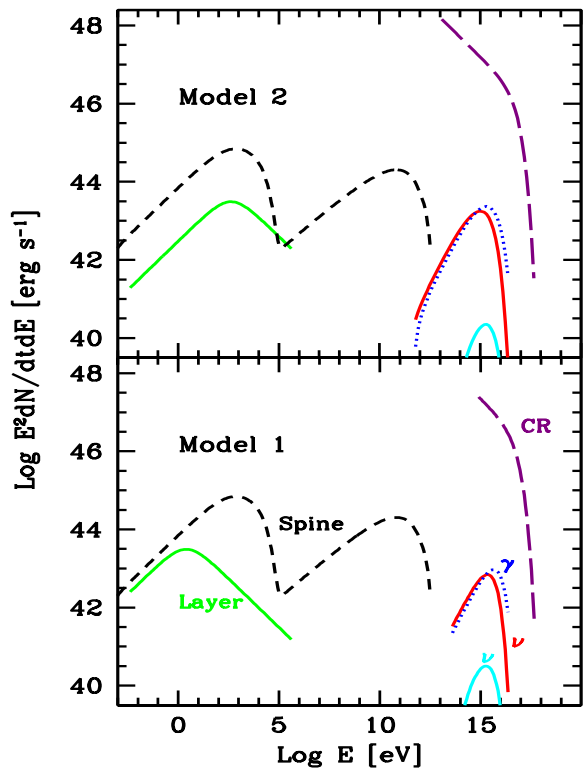}
\caption{\label{como} Luminosities of the different components in the observer frame for model 1 (bottom) and 2 (up). The solid green line shows the layer soft emission while, for comparison, the black dashed line is the blazar spectrum template (assumed to be emitted by the spine). The violet long-dashed line shows the spectrum of the high-energy protons. The solid red and the dotted blue lines show the luminosity of neutrinos (all flavors) and $\gamma$ rays produced through photo-meson reactions. For comparison, the cyan solid line shows the neutrino luminosity considering the internal synchrotron photons as targets for the photo-meson reaction.}
\end{figure}  

Summarizing, the free parameters of the model are:  the jet radius, $R$, the spine and layer Lorentz factors $\Gamma_{\rm s}$ and $\Gamma_{\rm l}$, the observed layer luminosity $L_{\rm t}$ and its peak energy $\epsilon_{\rm o}$, the spectral slopes $\alpha_1$ and $\alpha_2$, the spine comoving CR luminosity $L^{\prime}_p$, the CR power law index $n$, the minimum and the cut-off energy $E^{\prime}_{\rm min}$, $E^{\prime}_{\rm cut}$. 

The choice of the values of some parameters is guided by the results of the modeling of HBL emission. For definiteness we fix the jet radius to $R=10^{15}$ cm (e.g. Tavecchio et al. 2010), $\Gamma_{\rm s}=15$, $\Gamma_{\rm l}=2$ and $\delta_{\rm s}=20$. Consequently, $\delta_{\rm l}=3.7$ and $\Gamma_{\rm rel}=4$. The {\it observed} luminosity of the low-energy emission component of the layer is constrained  from above, since we demand that the {\it observed} SED of HBL is dominated by the spine. For the low and the high energy slope we assume the customary values $\alpha_1=0.5$ and $\alpha_2=1.5$. For the spine SED 
we adopt as a template  the SED of the prototypical HBL Mkn 421 calculated in Tavecchio et al. (2010). 

The remaining free parameters are $L_{\rm t}$, $\epsilon_{\rm o}$ - determining the layer target photon spectrum - and $L_p^{\prime}$, $E^{\prime}_{\rm min}$, $E^{\prime}_{\rm cut}$ and $n$ - specifying the CR spectrum. Adjusting these parameters we can find the best solution reproducing the IceCube measurements, reported in Fig.\ref{bkg} (red and gray data points, from Aartsen et al. 2014). We report two possible cases (model 1 and 2). For both  we assume $n=2.5$ and we assume that a CR flux level comparable to the observed one (since the neutrino luminosity is proportional to the product of the CR luminosity and the density of target photons, one could relax this assumption allowing a larger luminosity for the layer). The other parameters are listed in Table 1.  

\begin{table} 
\begin{center}
\begin{tabular}{cccccc}
\hline
\hline
Model & $L_{\rm t}$ & $\epsilon_{\rm o}$   & $L^{\prime}_{\rm p}$  & $E^{\prime}_{\rm min}$  &  $E^{\prime}_{\rm cut}$\\

& [erg s$^{-1}$] & [eV] &  [erg s$^{-1}$] & [eV]& [eV] \\
\hline  
1& $1.9\cdot10^{44}$&2.5   & $2.5\cdot10^{42}$& $2\cdot10^{12}$ & $2.3\cdot10^{15}$ \\
2& $1.9\cdot10^{44}$& 410  & $1.8\cdot10^{43}$& $3\cdot10^{10}$ & $2\cdot10^{15}$ \\
\hline
\hline
\end{tabular}                                                         
\vspace{0.3 cm}
\caption{Input parameters for the models shown in Fig.\ref{bkg}.}
\end{center}
\label{tab1}
\end{table}                                                                  

In a first case (model 1, dashed line in Fig.\ref{bkg}), we assume that the spectrum lies below the IceCube upper limit around 1 PeV - thus implying that the neutrinos with $E_{\nu}<1$ PeV belong to a separate component (see e.g. He et al. 2013).  This condition, together with the upper limit at 10 PeV, implies a quite narrow spectrum. In a second case (model 2, solid line) we relax this condition, allowing a single component to describe the entire spectrum. In both cases the neutrino flux is supposed to cut-off in correspondence to the upper limit at 10 PeV, constraint that fixes the  maximum CR energy.

The main difference between the two cases is the peak energy of the layer component, $\epsilon_{\rm o}$, and $E^{\prime}_{\rm min}$. The effect of these parameters can be understood recalling the relation linking the neutrino energy  to that of the parent CR, $E_{\nu}\simeq 0.05 E_{\rm p}$, and the threshold condition for pion production, $E^{\prime}_{p}\epsilon^{\prime}_{\rm t}\gtrsim \, m_{\pi}m_{p}c^4$. To increase the flux at low energy one has thus to decrease $E^{\prime}_{\rm min}$ (decreasing the energy of the produced neutrinos), increasing at the same time $\epsilon_{\rm o}$ (to satisfy the threshold). In turn, decreasing the minimum CR energy leads to the increase of the total CR power (dominated by the low energy particles). This explains the different parameters of the two cases.

We recall  that at energies below $\sim 300$ TeV one expects the possible contribution from a hard atmospheric prompt component, whose actual level is however still uncertain. The gray data points in Fig.\ref{bkg} display the extraterrestrial flux for an increased flux of the prompt component to the level of the 90\% CL limit (Aartsen et al. 2014). Model 2 is in full agreement with these high-background data.

Fig. 2  shows the luminosities in the observer frame of the different components for the two models. Besides the spine emission, the low energy emission of the layer, neutrinos photons and CR, we report for comparison the neutrino luminosity considering only the internal synchrotron emission. The large boost of the neutrino output of the system caused by the presence of the amplified layer radiation field is clearly visible. The ultra-high energy $\gamma$ rays from $\pi^0$ decay interacting with low energy photons start electromagnetic cascades in the jet. Since the luminosity of the resulting reprocessed component is small compared to that of the spine, we neglect it.

\section{Discussion}

Our calculations demonstrate that a velocity structure of BL Lac jets leads to an effective boosting of the neutrino emission with respect to the one-zone scenario. The IceCube measurements can be reproduced assuming a layer luminosity and a CR flux compatible with the observations. 

Some caveats are however in order.
The budget output of spine is strongly unbalanced toward CR. In fact, the beaming-corrected power in radiation is $P_{\rm rad}=L_{\rm rad}/\Gamma_{\rm s}^2= 2.4\cdot 10^{43}$ erg s$^{-1}$ for both models, while that in CR, $P_{\rm CR}=L^{\prime}_{\rm CR}\, \delta_{\rm s}^4/\Gamma _{\rm s}^2$ is $P_{\rm CR}=10^{45}$ erg s$^{-1}$ for model 1 and  $P_{\rm CR}=7.2\cdot 10^{45}$ erg s$^{-1}$ for model 2, implying a ratio $P_{\rm CR}/P_{\rm rad}=41$ and 300, respectively (similar to Murase et al. 2014). To sustain such a power, the {\it total jet power} $P_{\rm jet}$ must be larger (or at least equal) to that in the CR component. A similar result is derived comparing the CR emissivity of our model, which is of the order of several $10^{47}$ erg Mpc$^{-3}$ yr$^{-1}$, with the $\gamma$-ray emissivity of BL Lacs (e.g. Dermer \& Razzaque 2010) which is $<10^{47}$ erg Mpc$^{-3}$ yr$^{-1}$ and probably not exceeding $2-3\times 10^{46}$ erg Mpc$^{-3}$ yr$^{-1}$  for low-$z$ BL Lacs. Note that the ``curving proton" model of Dermer et al. (2014) should allow much smaller CR powers.

Another issue is related to the fact that, while we assume that CR accelerated in BL Lacs jets dominate  at $10^{16}$ eV, the standard view posits that their origin is galactic (e.g. Antoni et al. 2005). This difficulty could  be avoided if CR cannot efficiently escape from BL Lac jets/environment or outflowing winds prevent CR of these energies to penetrate into our Galaxy.

Considering the derived neutrino luminosity one can also calculate the expected number of events from a single source. Convolving the IceCube effective area with the flux derived with our model (as e.g. Yacobi et al. 2014) from a source at $z=0.03$ (like the prototypical HBL Mkn 421 and Mkn 501), we found that with 3 years of exposure one should detect 1 neutrino for model 1 and 4 neutrinos for model 2. Interestingly, this value is compatible with the findings of Padovani \& Resconi (2014). Also comparable is the gamma-ray/neutrino luminosity ratio for Mkn 421 in our Fig. 2 and their Fig. 1.

We stress that in this work we have considered only HBL, for which the existence of a layer is well assessed. The possible presence of a layer with similar properties in the jets of other BL Lac (IBL and LBL) or even in FSRQ could lead to an important contribution of these sources to the observed neutrino flux. A similar remark concerns the possible neutrino emission of weak (FRI) radiogalaxies, recently considered by Becker Tjus et al. (2014). In the blazar unification scheme, FRI radiogalaxies are the misaligned version of the BL Lac objects. In the framework of the structure jet model, the emission from their jets is expected to be dominated by the (slightly) Doppler boosted emission of the layer, since the spine flux is, at large enough $\theta _{\rm v}$, de-boosted by the relativistic Doppler effect. The amplification of the photo-meson luminosity induced by the spine-layer radiative interplay is expected to occur also for the layer - the spine radiation field being amplified in the layer frame. The luminosity emitted by a single radiogalaxy is expected to be quite lower than that of a blazars, but this is partly compensated by the (expected) larger number of  sources. Therefore one can speculate that the IceCube flux, similarly to the $\gamma$-ray background, could contain the contribution from both aligned (blazars) and misaligned sources (radiogalaxies).

\section*{Acknowledgments}
FT acknowledges  contribution from  grant PRIN--INAF--2011. D.G. is supported by a grant from the U.S. Israel Binational Science Foundation. We thank the referee for useful suggestions and P. Padovani for discussions.

\end{document}